\documentclass[traditabstract]{aa} 
\usepackage{graphicx}
\usepackage{url}
\usepackage{natbib}
\usepackage{amsmath}
\bibpunct{(}{)}{;}{a}{}{,}
\usepackage{txfonts}
\begin{document}
   \title{The O3N2 and N2 abundance indicators revisited: improved calibrations based on CALIFA and T$_{e}$-based literature data. }

   \subtitle{}

   \author{
   R.A.\,Marino \inst{1}
	\and F.F.\,Rosales-Ortega \inst{2,}\inst{3}
   	\and S.F.\,S\'{a}nchez \inst{4,}\inst{5}
	\and A.\,Gil de Paz \inst{1}
	\and J.\,V\'{i}lchez \inst{4}
	\and D.\,Miralles-Caballero \inst{2} 
	\and C.\,Kehrig \inst{4} 
	\and E.\,P\'{e}rez-Montero \inst{4} 
	\and V. Stanishev \inst{9}
	\and J.\,Iglesias-P\'{a}ramo \inst{4,}\inst{5}
	\and A.I.\,D\'{i}az \inst{2}
	\and A.\,Castillo-Morales \inst{1}
	\and R.\,Kennicutt \inst{6}
	\and A.R.\,L\'{o}pez-S\'{a}nchez \inst{7,}\inst{8}
	\and L.\,Galbany \inst{9}
	\and R.\,Garc\'{i}a-Benito \inst{4}
	\and D.\,Mast \inst{4,}\inst{5}
	\and J.\,Mendez-Abreu \inst{10,}\inst{11}
	\and A.\,Monreal-Ibero \inst{4}
	\and B.\,Husemann \inst{12}
	\and C.J.\,Walcher \inst{12}
	\and B.\,Garc\'{i}a-Lorenzo \inst{10,}\inst{11}
	\and J.\, Masegosa \inst{4}
	\and A.\, del Olmo Orozco \inst{4}
	\and A.M.\, Mour\~{a}o \inst{9}
	\and B.\,Ziegler \inst{13}
	\and M.\,Moll\'{a} \inst{14}
	\and P.\,Papaderos \inst{15}
	\and P.\,S\'{a}nchez-Bl\'{a}zquez \inst{2}
	\and R.M.\,Gonz\'{a}lez Delgado \inst{4} 
	\and J.\,Falc\'{o}n-Barroso \inst{11,}\inst{10}
	\and M.M.\,Roth \inst{12}
	\and G.\,van de Ven \inst{16}
     	\and the CALIFA team \inst{4}\fnmsep\thanks{Based on observations collected at the German-Spanish Astronomical Center, Calar Alto, jointly operated by the Max-Planck-Institut f\"{u}r Astronomie Heidelberg and the Instituto de Astrof\'{i}sica de Andaluc\'{i}a (CSIC).}
          }

   \institute{
   CEI Campus Moncloa, UCM-UPM, Departamento de Astrof\'{i}sica y CC$.$ de la Atm\'{o}sfera, Facultad de CC$.$ F\'{i}sicas, Universidad Complutense de Madrid, Avda.\,Complutense s/n, 28040 Madrid, Spain. \\
    \email{ramarino@ucm.es}
	\and 
	Departamento de F\'{i}sica Te\'{o}rica, Universidad Aut\'{o}noma de Madrid, 28049 Madrid, Spain.
	\and
	Instituto Nacional de Astrof{\'i}sica, {\'O}ptica y Electr{\'o}nica, Luis E. Erro 1, 72840 Tonantzintla, Puebla, Mexico.
	\and
	Instituto de Astrof\'{i}sica de Andaluc\'{i}a (CSIC), Camino Bajo de Hu\'{e}tor s/n, Aptdo. 3004, E18080-Granada, Spain.
	\and 
	Centro Astron\'{o}mico Hispano Alem\'{a}n, Calar Alto, (CSIC-MPG), C/Jes\'{u}s Durb\'{a}n Rem\'{o}n 2-2, E-04004 Almeria, Spain. 
	\and
         University of Cambridge, Institute of Astronomy Madingley Road, Cambridge, CB3 0HA, United Kingdom.
         \and
         Australian Astronomical Observatory, PO Box 915, North Ryde, NSW 1670, Australia.
         \and
         Department of Physics and Astronomy, Macquarie University, NSW 2109, Australia.
          \and
         CENTRA - Instituto Superior Tecnico, Av. Rovisco Pais, 1, 10 49-001 Lisbon, Portugal.
          \and
         Depto. Astrof\'{i}sica, Universidad de La Laguna (ULL), E-38206 La Laguna, Tenerife, Spain.
          \and
         Instituto de Astrof\'{i}sica de Canarias (IAC), E-38205 La Laguna,Tenerife, Spain.
         \and
         Leibniz-Institut f\"{u}r Astrophysik Potsdam (AIP), An der Sternwarte 16, D-14482 Potsdam, Germany.
         \and
         University of Vienna, Department of Astrophysics, T\"{u}rkenschanzstrasse 17, 1180, Vienna, Austria.
         \and 
         CIEMAT, Departamento de Investigaci\'{o}n B\'{a}sica, Avda. Complutense 40, 28040 Madrid, Spain.
         \and
         Centro de Astrof\'{i}sica and Faculdade de Ciencias, Universidade do Porto, Rua das Estrelas, 4150-762 Porto, Portugal. 
         \and
         Max Planck Institute for Astronomy, K\"onigstuhl 17, 69117 Heidelberg, Germany.}
   \date{}

 \abstract
 {The use of integral field spectroscopy is since recently allowing to measure the emission line fluxes of an increasingly large number of star-forming galaxies both locally and at high redshift. Many studies are using these fluxes to derive gas-phase metallicity of the galaxies applying the so-called strong-line methods. However, the metallicity indicators that these datasets use were empirically calibrated using few direct abundance data (T$_{e}$-based measurements). Furthermore, a precise determination of the prediction intervals of these indicators is commonly lacking in these calibrations. Such limitations might be leading to systematic errors in the determination of the gas-phase metallicity, especially at high redshift, that might have an important impact on our understanding of the chemical evolution of the Universe. The main goal of this study is to review the most widely used empirical oxygen calibrations, O3N2 and N2, by using new direct abundance measurements. We pay special attention to (1) the expected uncertainty of these calibrations as a function of the index value or abundance derived and (2) the presence of possible systematic offsets. This is possible thanks to the analysis of the most ambitious compilation of T$_{e}$-based \ion{H}{ii} regions to date. This new dataset compiles the T$_{e}$-based abundances of 603 \ion{H}{ii} regions extracted from the literature but also includes new measurements from the CALIFA survey. Besides providing new and improved empirical calibrations for the gas abundance, we also present here a comparison between our revisited calibrations with a total of 3423 additional CALIFA \ion{H}{ii} complexes with abundances derived using the ONS calibration by Pilyugin et al. (2010). The combined analysis of T$_{e}$-based and ONS abundances allows us to derive their most accurate calibration to date for both the O3N2 and N2 single-ratio indicators, in terms of all statistical significance, quality and coverage of the space of parameters. In particular, we infer that these indicators show shallower abundance dependencies and statistically-significant offsets compared to those of Pettini \& Pagel (2004), Nagao et al. (2006) and P\'{e}rez-Montero \& Contini (2009). The O3N2 and N2 indicators can be empirically applied to derive oxygen abundances calibrations from either direct abundance determinations with random errors of 0.18 and 0.16, respectively, or from indirect ones (but based on a large amount of data) reaching an average precision of 0.08 and 0.09 dex (random) and 0.02 and 0.08 dex (systematic; compared to the direct estimations), respectively. }

\titlerunning{The N2 and O3N2 abundance indicators revisited.}

\maketitle


\section{Introduction}

\ion{H}{ii} regions are a powerful tool to understand the physical properties and chemical evolution of the interstellar medium (ISM) in galaxies. They also represent the perfect laboratories for deriving chemical abundances of gaseous nebulae and stars clusters across the surface of nearby galaxies \citep[][hereafter OF06]{2006agna.book.....O}. From the study of their characteristic emission-line spectra we can get key insights not only on the amount of massive stars being formed and the rate star-formation occurs but also on nucleosynthesis processes taking place in these stars and on the physical conditions of the gas surrounding them \citep[see][for a review]{shields}.

With regard to the latter, the depth and spatial resolution of the spectroscopic observations carried out in the Local Universe have allowed not only to study the global (stars and gas) metallicity of galaxies but also the variation across galaxies and the abundance discrepancies among different phases of the ISM and between these and that of the stars. However, these measurements require the determination of the electron temperature (T$_{e}$), which is obtained from ratios of faint auroral to nebular emission line intensities, such as [O\,{\textsc{iii}}]$\lambda$4363/$\lambda$5007. This is often referred to as the {\it direct} (or T$_{e}$-based) method (OF06). A well-known difficulty, however, arises from the fact that auroral lines are inherently weak, and they become fainter as metallicity increases (as the temperature decreases, due to the more efficient cooling via metal lines). 
This has led to the need of using {\it indirect} abundances based on relations between metallicity and the (either predicted or empirically-calibrated) intensity of strong emission lines. This is particularly relevant when the issue of the chemical evolution of the Universe as a whole is addressed as only the brightest emission lines (from the brightest \ion{H}{ii} regions) are usually accessible at high-redshift \citep{pettini01, kobul, lilly, steidel}.

Therefore, calibrators based on strong emission lines are required in such cases. Most of these calibrators rely on comparison of the ratios between the strongest emission lines with direct estimations of the oxygen abundances \citep[e$.$g$.$][and references therein, hereafter P12]{p12}. A few of them rely alternatively on the comparison of these ratios with photoionization models \citep[e.g$.$][]{mcgaugh, 2002ApJS..142...35K}, which had the advantage of covering any possible range of parameters but carry our limitations in the knowledge on the actual physical conditions of the nebulae. 
These most widely used strong-line indicators are mainly focused on the relative intensity of the following lines: [\ion{O}{ii}]$\lambda$3727, H$\beta$, [\ion{O}{iii}]$\lambda$5007, H$\alpha$, [\ion{N}{ii}]$\lambda$6583, [\ion{S}{ii}]$\lambda\lambda$6717,6731. \cite[][hereafter LS12]{angel12} have recently presented a revision on the different methods, showing their main strengths and caveats and illustrating in which range/conditions they can be safely applied.
The goal of our work is to provide updated calibrations for two (O3N2 and N2) of the most widely used indicators of the oxygen abundance. These indicators are very popular at low and high redshift for different reasons: (1) O3N2 is weakly affected (compared to the R$_{23}$ or [\ion{N}{ii}]$\lambda$6583/[\ion{O}{ii}]$\lambda$3727 indexes) by differential extinction and makes use of the strongest and most easily accessible emission lines in the rest-frame optical spectroscopy, (2) N2 uses two emission lines that are very close in wavelength and are also accessible in the near-infrared at moderate-to-high redshifts \citep[see works by ][]{cresci, sinfoni}. However, N2 does not consider an ionization parameter of the gas, which may be very important in some cases, specially when dealing with integral field spectroscopy (IFS) data \citep[][hereafter LS11;  \citealt{carol13}]{angel11}. The most popular calibration of the O3N2 and N2 indicators was introduced by \citet{pp04} (hereafter PP04). Their calibration has the fundamental problem that it combined direct derivations of the oxygen abundance for the metal-poor range, with estimations based on photoionization models for the high-metallicity range. This is prompt to large uncertainties due to the well-known difference in the absolute scale of abundances between both methods \citep[][]{kewell}. For the sake of completeness, we will also compare our calibrations with the ones proposed by \cite{nag06} (hereafter NAG06) and \cite{pmc09} (hereafter PMC09) that are based, respectively, on Sloan Digital Sky Survey data \citep[SDSS,][]{york00,strauss02} and literature measurements.

In this work we address the re-calibration of both indicators by means of anchoring them to direct estimations of the oxygen abundance. In order to do so we take advantage of two complementary datasets. First, we use the recent compilations of direct abundance measurements by P12 and PMC09, together with other recently published values.  We also present here, for the first time, new T$_{e}$-based measurements belonging from the CALIFA survey (16 \ion{H}{ii} regions with [O\,{\textsc{iii}}]$\lambda$4363\,). This compilation led to a notorious increase in the number of regions with direct abundance measurements even at the elusive high-metallicity range. For the whole T$_{e}$-sample, due to the heterogeneous nature of our compilation, we have re-calculated the electron temperatures and the oxygen abundances of the 603 \ion{H}{ii} regions using the recipe proposed by \cite{p10} (hereafter P10) and P12 to homogenize our sample. 

Secondly, we use the large catalog of extragalactic \ion{H}{ii} regions created by the CALIFA survey \citep[see][]{seba12, seba13}, that comprises emission-line flux measurements for thousands of \ion{H}{ii} complexes. The limits set to the CALIFA sample in terms of diameter and distance (see Sect$.$~2.2) prevents auroral lines to be detected except for the case of [O\,{\textsc{iii}}]$\lambda$4363\ and only in the nearest and most metal-poor galaxies in the sample. However, this dataset provides with unprecedented statistics in the measurement of multiple strong emission-line fluxes and their corresponding line ratios.

Thus, our objectives are to provide i) an accurate derivation of the oxygen abundance from the single-ratio indicators O3N2 and N2 anchored to {\it direct} abundance measurements. However, due to the still somewhat limited number of regions available and poor coverage in metallicity, it will have a limited capacity with regard of the estimation of the uncertainties at specific values of the aforementioned O3N2 and N2 indices. In order to partly mitigate this problem and combine with the results of the previous analysis, we carry out ii) a comparison between the new T$_{e}$-calibrations found with the best fit obtained using $\sim$3400 \ion{H}{ii} complexes from the CALIFA survey \citep{seba12} where oxygen abundances could be indirectly computed using the multiple line-ratio ONS calibration (P10). Thus, we will use the large collection of CALIFA \ion{H}{ii} regions to analyze not only the most probable value of the oxygen abundance for a given value of these single-parameter indicators but also the corresponding uncertainty and in the widest range of physical parameters possible. This is particularly critical when the evolution of the metallicity of galaxies with redshift ought to be analyzed as any potential evolutionary effect should be compared with the individual uncertainties achieved at each specific redshift. In this regard, it should be emphasized that these are largely affected by the uncertainties associated with the calibration of the indices and the variation in the prediction interval with the index itself. The {\it indirect} derivations of the oxygen abundances analyzed in this work are T$_{e}$-anchored calibrations making use of multiple strong-line ratios. We will make use of these indirect metallicities of the CALIFA sample in the knowledge that some offsets between them could still be present due to the fact that the CALIFA sample is mapping a larger universe of physical conditions with respect to our T$_{e}$ sample.

The content of the article is distributed as follows: Sect$.$~2 describes the sample of \ion{H}{ii} regions, including those with T$_{e}$-based abundances in the literature and those in the CALIFA survey. In Sect$.$~3 we give details on the analysis procedures and our main results while Sect$.$~4 summarizes the conclusions of this work. 

\section{The sample}

This work is based on the largest accessible database of \ion{H}{ii} regions ever accomplished, including a compilation of 603 \ion{H}{ii} regions with accurate measurements of the electron temperature, together with 3423 \ion{H}{ii} regions provided by the CALIFA survey \citep{seba13}. As we describe below, most of the regions in the CALIFA catalog (all except 16) could only be used to understand the behavior pattern of the different single-line ratio estimators compared to {\it indirect} abundance measurements. On the other hand, the emission-line data from the literature (plus 16 CALIFA \ion{H}{ii} complexes) are used to determine {\it direct} oxygen abundances for the empirical calibrations of O3N2 and N2 indices. 

\subsection{Compilation of T$_{e}$-based \ion{H}{ii} regions}

We have performed a comprehensive search in the literature for \ion{H}{ii} regions within spiral and irregular galaxies in order to compile our T$_{e}$-based sample. We have looked for those targets in previous work having measurements of bright emission lines (typically [O\,{\textsc{iii}}]$\lambda$5007 and [N\,{\textsc{ii}}]$\lambda$6584) and at least one of the auroral emission line [O\,{\textsc{iii}}]\,$\lambda$\,4363, [N\,{\textsc{ii}}]\,$\lambda$\,5755, [S\,{\textsc{iii}}]\,$\lambda$\,6312, because we want to recalculate all the temperatures and the indices in an homogeneous way. The compilation consists of a set of 603 calibrating \ion{H}{ii} regions from 17 different works in the literature whose references are given in Table~\ref{Sample}. In our compilation we eliminated \ion{H}{ii} regions that were found to be duplicated among different works that actually came from the same original observational dataset. For this reason we have selected only 84 \ion{H}{ii} regions from the work of PMC09 as the ones not in common with the P12 sample.

\begin{table}[]
\caption{Bibliographic references to the original works for the compiled T$_{e}$-sample.}\label{Sample}
\begin{center}
\begin{tabular}{l c c}
\hline\hline
Reference &  Number of   & Auroral lines$^{a}$\\
 & \ion{H}{ii} regions & \\
\hline
\hline
\cite{berg12} & 2 & [O\,{\textsc{iii}}]\\
\cite{bres12}  & 16 & [O\,{\textsc{iii}}]\\
\cite{crow09}  & 4 & [O\,{\textsc{iii}}]\\
\cite{crox09} & 2 & [O\,{\textsc{iii}}]\\
\cite{esteban13} & 1 & [O\,{\textsc{iii}}]\\
\cite{ruben} & 3 & [O\,{\textsc{iii}}], [S\,{\textsc{iii}}]\\
\cite{gus12} & 3 & [O\,{\textsc{iii}}]\\
\cite{had07} & 6 & [O\,{\textsc{iii}}]\\
\cite{keh11} & 3 & [O\,{\textsc{iii}}]\\
\cite{monreal12} & 3 & [O\,{\textsc{iii}}]\\
\cite{pmc09} & 84 & [O\,{\textsc{iii}}], [N\,{\textsc{ii}}]\\
\cite{p12} & 414 & [O\,{\textsc{iii}}], [N\,{\textsc{ii}}], [S\,{\textsc{iii}}]\\
\cite{sand12} & 5 & [O\,{\textsc{iii}}]\\
\cite{sta13} & 16 & [O\,{\textsc{iii}}]\\
\cite{wes} & 7 & [N\,{\textsc{ii}}]\\
\cite{zah11} & 9 & [O\,{\textsc{iii}}]\\
\cite{zur12} & 9 & [O\,{\textsc{iii}}], [N\,{\textsc{ii}}], [S\,{\textsc{iii}}]\\
{\bf  This work}& {\bf 16} & {\bf [O\,{\textsc{iii}}]}\\
\hline 
{\bf Total} & {\bf 603} & \\
\hline\hline
\end{tabular}
\end{center}
$^{a}$ The corresponding wavelengths of the auroral lines are [O\,{\textsc{iii}}]\,$\lambda$\,4363, [N\,{\textsc{ii}}]\,$\lambda$\,5755, [S\,{\textsc{iii}}]\,$\lambda$\,6312. \\
\end{table}

\subsection{The CALIFA catalog of \ion{H}{ii}  Regions}

\begin{table*}[t]
\caption{Dereddened emission line ratios relative to $\mathrm{H}\beta$ fluxes of the CALIFA T$_{e}$ sample along with their errors.}\label{fluxes4363}
\begin{center}
\begin{tabular}{l c c c c c c}  
\hline\hline
ID & [O\,{\textsc{ii}}]$\lambda$\,3726,29 & [O\,{\textsc{iii}}]$\lambda$\,4363 & [O\,{\textsc{iii}}]$\lambda$\,5007 &  [N\,{\textsc{ii}}]$\lambda$\,6583 & F$_{\mathrm{H}\beta}$ & A$_{V}$ \\
\hline\hline
MCG-01-54-016-001 & 3.83$\pm$0.06 & 0.041$\pm$0.006 & 3.11$\pm$0.05 &  0.12$\pm$0.02 & 139.78$\pm$0.34 & 0.63\\
NGC3991-001 & 2.88$\pm$0.12 & 0.018$\pm$0.009 & 2.73$\pm$0.11 &  0.24$\pm$0.04 & 1859.3$\pm$2.9 & 0.58\\
NGC3991-002 & 3.98$\pm$0.11 & 0.016$\pm$0.007 & 2.18$\pm$0.08 &  0.30$\pm$0.04 & 1019.1$\pm$1.7 & 0.68\\
NGC3991-007 & 5.18$\pm$0.07 & 0.013$\pm$0.004 & 1.77$\pm$0.04 &  0.37$\pm$0.03 & 233.58$\pm$0.46 & 1.30\\
NGC7489-002 & 3.35$\pm$0.09 & 0.043$\pm$0.010 & 3.19$\pm$0.07 &  0.26$\pm$0.02 & 87.13$\pm$0.37 & 0.88\\
NGC7489-007 & 3.87$\pm$0.13 & 0.080$\pm$0.016 & 3.87$\pm$0.10 &  0.26$\pm$0.03 & 80.43$\pm$0.47 & 0.98\\
UGC00312-001 & 3.51$\pm$0.04 & 0.015$\pm$0.003 & 2.93$\pm$0.04 & 0.23$\pm$0.02 & 512.98$\pm$0.53 & 1.36\\
UGC00312-004 & 3.07$\pm$0.04 & 0.016$\pm$0.003 & 2.68$\pm$0.04 & 0.25$\pm$0.02 & 296.63$\pm$0.39 & 0.91\\
UGC00312-005 & 3.97$\pm$0.04 & 0.018$\pm$0.003 & 2.38$\pm$0.03 & 0.26$\pm$0.02 & 155.21$\pm$0.24 & 0.90\\
UGC00312-007 & 3.65$\pm$0.04 & 0.018$\pm$0.003 & 2.82$\pm$0.03 & 0.24$\pm$0.02 & 155.22$\pm$0.24 & 0.99\\
UGC00312-008 & 3.00$\pm$0.05 & 0.034$\pm$0.005 & 3.12$\pm$0.04 & 0.19$\pm$0.02 & 59.91$\pm$0.17 & 0.60\\
UGC00312-013 & 4.33$\pm$0.06 & 0.033$\pm$0.006 & 1.92$\pm$0.03 & 0.29$\pm$0.02 & 39.53$\pm$0.13 & 0.91\\
UGC08733-004 & 2.44$\pm$0.05 & 0.021$\pm$0.006 & 2.25$\pm$0.04 & 0.26$\pm$0.03 & 46.46$\pm$0.16 & 0.68 \\
UGC10331-004 & 5.78$\pm$0.07 & 0.019$\pm$0.004 & 2.12$\pm$0.04 & 0.39$\pm$0.02 & 98.87$\pm$0.26 & 1.79\\
UGC10796-001 & 3.14$\pm$0.04 & 0.032$\pm$0.004 & 1.91$\pm$0.03 & 0.32$\pm$0.03 & 94.54$\pm$0.18 & 0.84\\
UGC12494-001 & 2.51$\pm$0.11 & 0.064$\pm$0.015 & 4.67$\pm$0.12 & 0.12$\pm$0.03 & 74.13$\pm$0.47 & 0.25\\
\hline\hline
\end{tabular}
\end{center}
NOTE: Fluxes are measured in units of $10^{-16}$\,erg\,cm$^{-2}$\,s$^{-1}$. The observed lines are dereddened with a \cite{1989ApJ...345..245C} extinction law. A$_{V}$ is calculated from the Balmer decrement ($\mathrm{H}\alpha$/$\mathrm{H}\beta$) with adopting a MW extinction law (A$_{V}$/A$_{H_{\beta}}$ =1.164, \cite{1989ApJ...345..245C}).
\end{table*}

The Calar Alto Legacy Integral Field Area (CALIFA) project is one of the most ambitious 2D-spectroscopic surveys to date, which will provide to the scientific community a very large number of spectra of individual \ion{H}{ii} regions in nearby galaxies. This ongoing large project will use 250 observing nights already awarded with the Centro Astron\'{o}mico Hispano Alem\'{a}n (CAHA) 3.5m telescope. CALIFA comprises a diameter (45" $<$ D$_{25}$ $<$80") selected sample of $\sim$600 galaxies in the Local Universe (0.005 $<$ z $<$ 0.03) addressing several fundamental issues in galactic structure and evolution of galaxies.

The CALIFA observations started in July 2010 and are performed using the Potsdam Multi Aperture Spectrograph, PMAS \citep{roth} at CAHA 3.5m in the PPAK mode \citep{ver, 2006PASP..118..129K}. The PPAK Integral Field Unit (IFU) has an hexagonal field-of-view (FoV) of $74\arcsec \times 62 \arcsec$, sufficient to cover the full optical extent of the galaxies up to 2-3 effective radii, in average. This is possible due to the diameter selection of the sample (Walcher et al., in prep.). PPAK is made up of 331 science fibers, 36 sky background fibers and 15 calibration fibers with a diameter of 2.7$\arcsec$ and has a 100\% covering factor when a three-pointing dithering scheme is adopted. The CALIFA spectra allow us to investigate the most prominent emission lines from our \ion{H}{ii} regions covering the wavelength range of 3700-7300\,\AA\AA\ in two overlapping setups: V500 (from 3745\,\AA\, to 7300\,\AA\, with a resolution of $\sim$ 850) and V1200 (from 3700\,\AA\, to 4750\,\AA\, with a resolution of $\sim$ 1650) as showed in Fig$.$~\ref{spectra}. 

The CALIFA data used for this work are based on the products generated by the CALIFA pipeline \citep[version 1.3c, see][]{bernd13}. Our data fulfill the foreseen quality control requirements, with a spectrophotometric accuracy better than 10\%\ everywhere within the considered wavelength range, both absolute and relative, and a depth that allows us to detect emission lines in individual \ion{H}{ii} regions as weak as $\sim$10$^{-17}$\,erg\,s$^{-1}$\,cm$^{-2}$\, with a S/N$\sim$3-5. For more details on the sample selection, observation strategy, data reduction procedures implemented in the pipeline and data quality control the reader is referred to \cite{seba12a}. The galaxies under study have been selected from the CALIFA observed sample. Since CALIFA is an ongoing survey, whose observations are scheduled on a monthly basis (i.e., dark nights), the list of objects increases regularly. The current results are based on the 150 galaxies observed until July 2012.

We used \textsc{HIIexplorer}\footnote{\textsc{HIIexplorer} web page: \url{http://www.caha.es/sanchez/HII_explorer/}} to extract the spectroscopic properties of our CALIFA catalog of \ion{H}{ii} regions sample from the V500-setup datacubes. \textsc{HIIexplorer} is an automatic \ion{H}{ii} region detection code described in \cite{seba12} that allow us to select \ion{H}{ii}  regions based on the contrast of the H$\alpha$ line intensity. This catalog has been recently used to study the mass-metallicity relation, and the possible dependence with the star formation rate \citep{seba13}. In a companion article we will describe the procedure for this particular dataset, summarizing the main properties of the ionized regions across these galaxies (S\'{a}nchez et al., in prep.). We present here just a brief summary of the different steps included in the overall process: first we create a narrow-band image of 120\,$\AA$ width, centered at the wavelength of H$\alpha$ shifted at the redshift of the targets. The narrow-band image is properly corrected for the contamination of the adjacent continuum; the narrow band image is used as an input for \textsc{HIIexplorer}. The code provides with a segmentation map that identifies each detected ionized region. Then, it extracts the integrated spectra corresponding to each segmented region. 

A total of 4942 individual \ion{H}{ii} regions are selected from the data-cubes of 150 galaxies, most of them part of the 1st CALIFA Data Release \citep[][]{bernd13} with the rest being part of the whole CALIFA mother sample (Walcher et al., in prep.). This sample comprises galaxies of any type, mostly spirals (both early and late type), with and without bars, and with different inclinations. In addition to the extraction of the V500 spectra, on which the analysis of the indirect metallicity measurements from the CALIFA data is based, we have also extracted the V1200 spectra for those same regions and segmentation maps in order to improve the detectability of the weak [O\,{\textsc{iii}}]\,$\lambda$\,4363\, line in the case of the T$_{e}$-based abundance measurements. In order to understand the physical properties of the \ion{H}{ii} regions we need to investigate only the emission features in our spectra, for this reason (i) each extracted spectrum was decontaminated by the underlying stellar continuum using the multi-SSP (Simple Stellar Populations) fitting routines included in FIT3D software tool \citep[][]{sebafit3d1, sebafit3d2}; (ii) then each emission line within the considered wavelength range was fitted with a Gaussian function to determine the line intensities and ratios; (iii) these line ratios were used to discriminate between different ionization conditions and to derive the oxygen abundances for each particular \ion{H}{ii} region/complex.

Finally, out of the $\sim$5000 \ion{H}{ii} regions detected with \textsc{HIIexplorer}, we kept a total of 3423 regions where a blue underlying continuum was clearly detected. This allows us to exclude regions where the ionized-gas emission is not associated with massive star formation. Quantitatively speaking we imposed that at least 20\% of the light at 5000\,\AA\ was arising from a population younger than 500 Myr, according to the multi-SSP spectral fitting provided by FIT3D \citep[see more details in][]{cid, seba13}. Although this limit could also exclude low-burst-strength regions in the bulges of our galaxies it will ensure that only ionized-gas emission induced by massive star formation is considered hereafter i$.$e$.$, it is a conservative limit.
 
Once the \ion{H}{ii} regions are selected, all the stellar-decontaminated spectra are analyzed and a final catalog of their spectroscopic properties is created for each galaxy. This catalog\footnote{An example of a similar catalog is described in \cite{sebacat}.} comprises the flux intensity and estimated error for the most prominent emission lines within the considered wavelength range, following the scheme presented in the Appendix 1 of \cite{seba12}, and includes [O\,{\textsc{ii}}]\,$\lambda$\,3727;  [O\,{\textsc{iii}}]\,$\lambda$\,4363\,; [O\,{\textsc{iii}}]\,$\lambda\lambda$\,4959,5007; $\mathrm{H}\beta$; [N\,{\textsc{ii}}]\,$\lambda$\,6548; $\mathrm{H}\alpha$; [N\,{\textsc{ii}}]\,$\lambda$\,6583\ and [S\,{\textsc{ii}}]\,$\lambda\lambda$\,6717,6731. Finally, we also corrected the emission fluxes for interstellar reddening using a value of 2.86 for the theoretical $\mathrm{H}\alpha$/$\mathrm{H}\beta$ ratio, assuming Case B with electron temperature $\sim$10,000 K and density $\sim$100 cm$^{-3}$ (OF06).

\subsection{The CALIFA T$_{e}$-based \ion{H}{ii} Regions}

\begin{figure*}[t]
   \centering
  \includegraphics[scale=0.45]{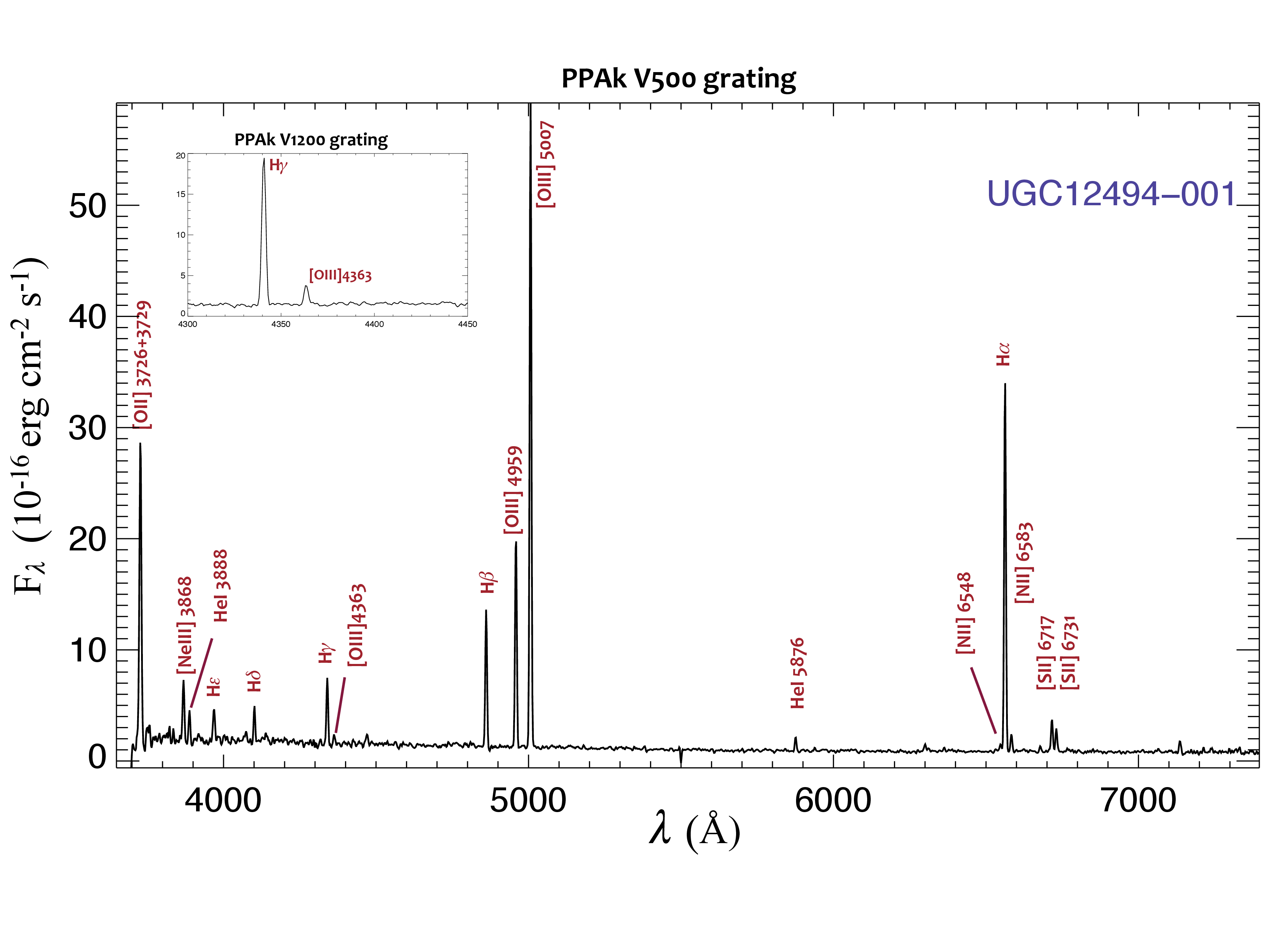}
   \caption{Representative PPAK V500 (main window) and V1200 (inset window) gratings optical spectra for the \ion{H}{ii} region UGC12494-001. The brightest optical emission lines of the CALIFA V500 spectral range are labeled. In the inset window a section of V1200 spectrum around the [O\,{\textsc{iii}}]\,$\lambda$\,4363 line (used to compute the electron temperature) is showed. For both gratings the fluxes are in units of $10^{-16}$\,erg\,cm$^{-2}$\,s$^{-1}$.  }
              \label{spectra}%
    \end{figure*}

As commented above, in addition to this catalog, which is based on the CALIFA V500 data, we also analyzed the V1200 extracted spectra for those regions where the [O\,{\textsc{iii}}]$\lambda$4363 line was tentatively detected. The high spectral resolution data from CALIFA are important for improving the detection rate of weak emission lines, such as this one, as they are strongly affected by spectral beam dilution. After imposing a cut on S/N ($>$ 3$\sigma$) at the peak of this line, a relative line flux error $\leq$ 20\% and visually inspecting the spectra for potential spurious detections, we identified a total of 16 CALIFA \ion{H}{ii} complexes where we can reliably derive O$^{++}$ zone electron temperatures and from these, using the formulation of P10, the oxygen abundance. In Table~\ref{fluxes4363} we present the most important emission line fluxes measured (relative to the $\mathrm{H}\beta$ fluxes) for our CALIFA-T$_{e}$ \ion{H}{ii} regions. The optical spectra of one of these targets (UGC12494-001) is shown in Fig$.$~\ref{spectra} where we are plotting the V500 grating in the main window and a section of the V1200 grating in the inset panel.

As extensively described in Sect$.$~1, our aim is to use the sample listed above to calibrate the most widely used indicators (O3N2, N2) versus (O/H) using both {\it direct} (bona fide calibrators that despite the dramatic improvement compared to previous calibration efforts are still somewhat limited in coverage and statistics) and {\it indirect} (based on multiple strong-line ratios but with extensive metallicity coverage and good statistics) methods to derive the gas metallicities. The following sections describe the details of the analysis and methods used along with our results. 
 
\begin{figure*}[t]
   \centering
\includegraphics[scale=1.0]{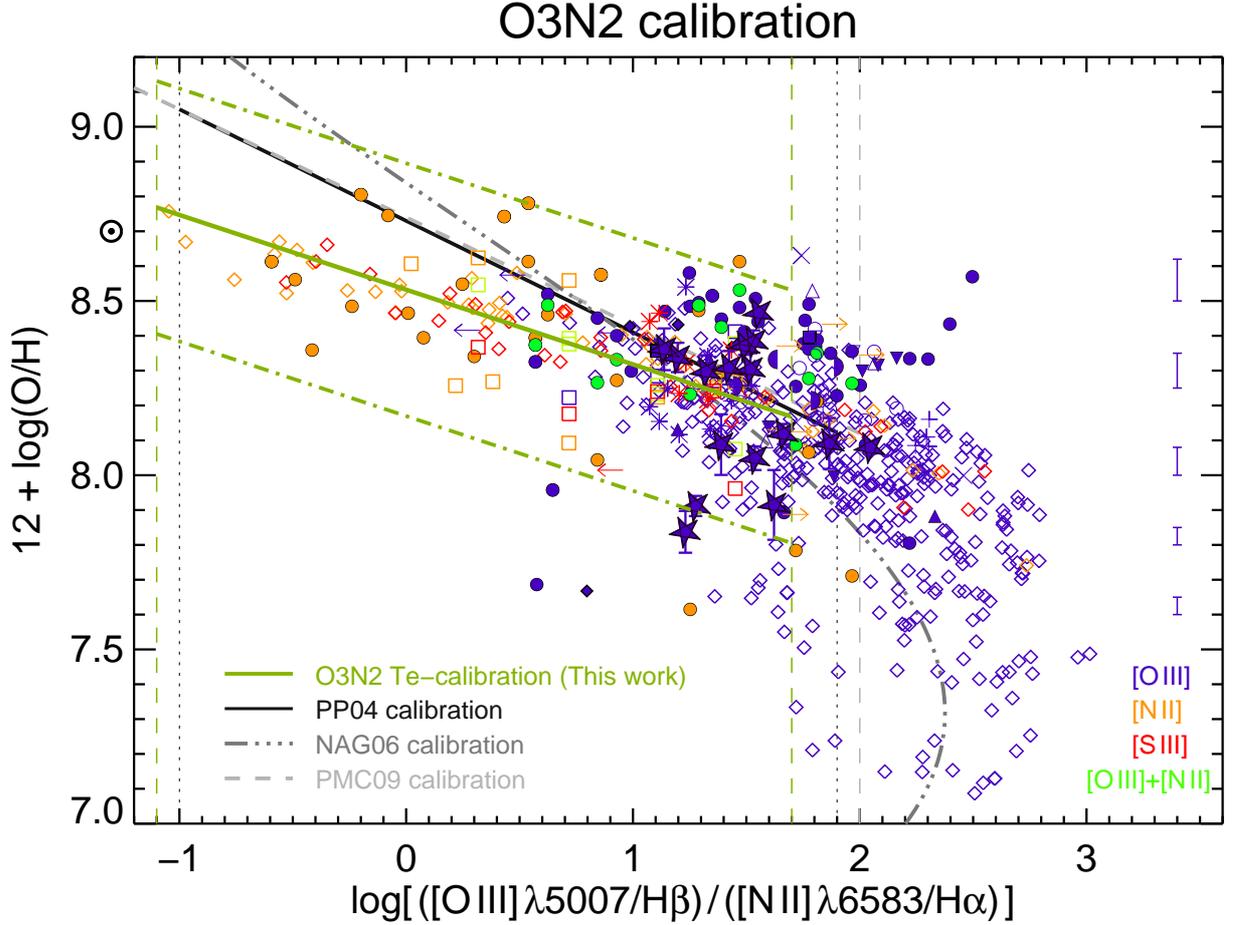}
   \caption{ Oxygen abundance versus the {\it O3N2} index for T$_{e}$-based \ion{H}{ii} regions abundances. The T$_{e}$-based \ion{H}{ii} regions are shown using different symbols and colors depending on the work they came from and the auroral line used for computing the electron temperature. The meaning of the symbols is the following: filled squares for \cite{berg12} data, empty circles for \cite{bres12} data, filled stars for the new CALIFA-T$_{e}$ data, plus signs for \cite{crow09} data, empty upside down triangles for \cite{crox09} data, filled triangles for \cite{gus12} data, filled upside-down triangles for \cite{had07} data, crosses for \cite{keh11} data, filled circles for \cite{pmc09} data, empty diamonds for \cite{p12} data, filled diamonds for \cite{sand12} data, asterisks for \cite{sta13} data, empty triangles for \cite{zah11} data, empty squares for \cite{zur12} data, left pointing arrows for \cite{ruben} data, right pointing arrows for \cite{wes} data, filled left semi-circles for \cite{esteban13} data and filled right semi-circles for \cite{monreal12} data. In blue we show those regions for which the temperature t$_{\,3,O}$ was computed using the [O\,{\textsc{iii}}]\,$\lambda$\,4363\, line, in orange are those regions for which we obtain the temperature t$_{\,2,N}$ from the [N\,{\textsc{ii}}]\,$\lambda$\,5755\, line, in red we represent those regions for which we are able to compute the temperature t$_{\,3,S}$ using the [S\,{\textsc{iii}}]\,$\lambda$\,6312, and in green are shown those regions that have both [O\,{\textsc{iii}}]\,$\lambda$\,4363\, and [N\,{\textsc{ii}}]\,$\lambda$\,5755\, measurements so we could compute the abundance using t$_{\,\,3,O}$ and t$_{\,2,N}$. The PP04 calibration is plotted with a black solid line with its applicability interval (from $-$1 to 1.9) shown with the vertical grey dotted lines. The PMC09 calibration is represented with a grey dashed line until its limit of validity O3N2 = 2. The grey 3dot-dashed line shows the NAG06 calibration. The new T$_{e}$-based calibration, 12 + log(O/H) = 8.533 - 0.214$\times$O3N2 is showed with a green solid line. The green dot-dashed lines encompass 2\,$\sigma$ (= $\pm$0.36 dex) of all measurements. The vertical dashed lines indicate the interval of our fit, from O3N2=$-$1.1 to 1.7. We only plot in blue the errors associated to the new measurements of the CALIFA-T$_{e}$ \ion{H}{ii} regions. For the sake of clarity we show in the right the average random error associated to the computation of the oxygen abundances for different abundance bins.}
              \label{o3n2_te}%
    \end{figure*}
    
    \begin{figure*}[t]
   \centering
\includegraphics[scale=1.0]{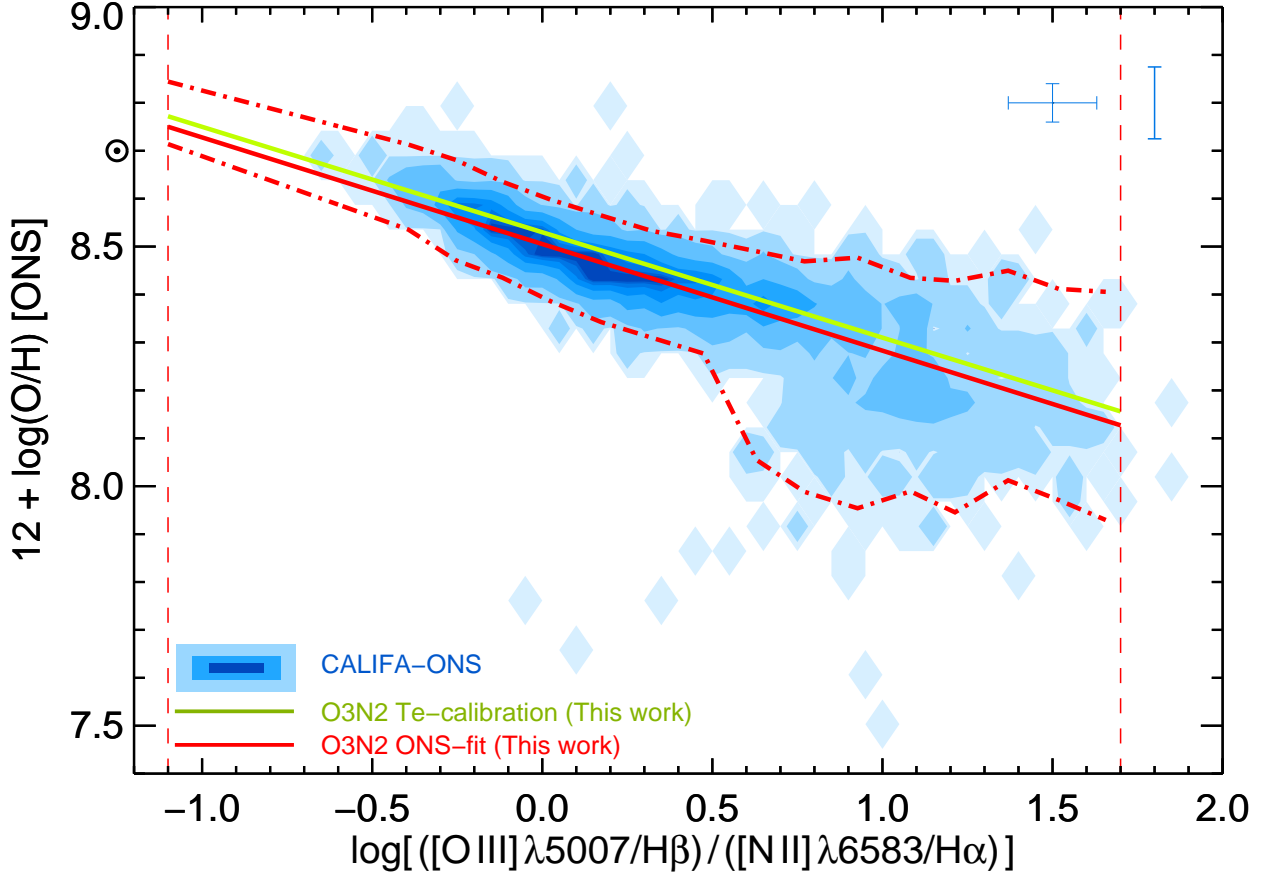}
   \caption{Oxygen abundance versus the {\it O3N2} index for the \ion{H}{ii} complexes detected within the CALIFA galaxies. For the sake of clarity we plot the CALIFA-ONS \ion{H}{ii} regions as a blue density contour-plot. Our new O3N2 calibration (see Fig$.$2) is plotted with a green solid line. Red lines are used to show the CALIFA-ONS best linear fit (solid) and its 2\,$\sigma$ prediction intervals (dot-dashed) with its applicability interval (from $-$1.1 to 1.7, vertical dashed lines) (see text for details on these fits). In this case we obtain 12 + log(O/H) = 8.505 - 0.221$\times$O3N2. The blue cross at the top right corner of the plot indicate the average random (left) of the CALIFA-ONS data points and the typical systematic (right) error associated to the measurement of ionized-gas oxygen abundances. }
              \label{o3n2_califa}%
    \end{figure*}

\section{Analysis and Results}

\begin{table}[]
\caption{Temperatures and oxygen abundances derived for the CALIFA T$_{e}$ sample using the [O\,{\textsc{iii}}]$\lambda$\,4363\, auroral line.}\label{results4363}
\begin{center}
\begin{tabular}{l c c c }  
\hline\hline
ID & S/N$_{\rm[O\,{\textsc{iii}}]\lambda\,4363 }$ & T$_{3}$ [K] & 12+log(O/H) \\
\hline\hline
MCG-01-54-016-001 & 8.2 &  12750$\pm$70   &  8.09$\pm$0.07 \\
NGC3991-001 & 10.7 &  10070$\pm$330 &  8.30$\pm$0.05  \\
NGC3991-002  & 4.9 &  10430$\pm$360  &  8.30$\pm$0.05   \\
NGC3991-007  & 3.2 &  10340$\pm$460  &  8.36$\pm$0.06  \\
NGC7489-002 & 4.0 &  12920$\pm$330  &  8.05$\pm$0.03  \\
NGC7489-007 & 4.1 &  15470$\pm$930  &  7.92$\pm$0.10  \\
UGC00312-001 & 8.5 &  9360$\pm$120  &  8.47$\pm$0.02 \\
UGC00312-004  & 7.5 &  9670$\pm$180   &  8.37$\pm$ 0.03  \\
UGC00312-005 & 5.6 &  10450$\pm$370   &  8.31$\pm$0.05  \\
UGC00312-007 & 5.6 &  9940$\pm$240   &  8.38$\pm$0.03  \\
UGC00312-008 & 5.7 &  11920$\pm$230   &  8.12$\pm$0.03   \\
UGC00312-013  & 6.9 &  14340$\pm$200  &  7.91$\pm$0.03   \\
UGC08733-004 &  4.0 &  11160$\pm$710  &  8.09$\pm$0.09  \\
UGC10331-004 & 3.3 &  10990$\pm$340   &  8.34$\pm$0.04 \\
UGC10796-001 & 6.2 &  14090$\pm$930   &  7.84$\pm$0.06   \\
UGC12494-001 & 16.4 &  12970$\pm$120   &  8.08$\pm$0.02  \\
\hline\hline
\end{tabular}
\end{center}
\end{table}

A number of relations have been proposed in the literature to derive metal abundances and temperatures from the metallicity-sensitive emission lines ratios \citep[e.g.][]{dopeva, 1994ApJ...420...87Z, vilchez96, dpm00, p00, p01a, pp04, tremonti, pilythuan, liang, sta06, epm07, thuan10}. In this regard, we highlight the recent comparative study of LS12 and similar previous studies from \cite{pmd05, kewell} and \cite{angel10} (hereafter LSE10). Oxygen abundances anchored to T$_{e}$-based calibrations have been derived using different methods for each of the two types of data described in Sect$.$~2.1 and 2.2. For this reason we have investigated which kind of {\it indirect} empirical calibrators based on single line ratios could be most accurate in deriving the metallicity of \ion{H}{ii} regions when only shallow spectra (of high redshift galaxies, for example) are available. We also compute oxygen abundance using both {\it direct} (T$_{e}$-based) methods and different indirect derivations of the metallicity given in the literature and use the strength of both methods to best characterize the applicability and associated uncertainties of single strong-line ratios to infer oxygen abundances. On one hand, the electron temperatures and the metallicity have been computed from relations similar to the ones proposed by \cite{camp86}, \cite{pmd03} and \cite{p07} and the T$_{e}$-method (see e.g. P10). In particular we recalculate for all the targets of our T$_{e}$-sample the oxygen abundance using the equations proposed in Sec$.$2.2 of P10 and of P12. By combining the auroral and the nebular lines we are able to compute confident electron temperatures t$_{3,O}$ from the [O\,{\textsc{iii}}]\,$\lambda\lambda$\,4959+5007/[O\,{\textsc{iii}}]\,$\lambda$\,4363\, ratio, t$_{2,N}$ from [N\,{\textsc{ii}}]\,$\lambda\lambda$\,6548+6584/[N\,{\textsc{ii}}]\,$\lambda$\,5755\, ratio and t$_{2,O}$ using the [O\,{\textsc{ii}}]\,$\lambda$,3727/[O\,{\textsc{ii}}]\,$\lambda\lambda$\,7320+7330\, ratio that lead to oxygen abundances with a typical error of 0.12 dex. The electron temperature t$_{3,S}$ can be estimated from the [S\,{\textsc{iii}}]\,$\lambda\lambda$\,9069+9532/[S\,{\textsc{iii}}]\,$\lambda$\,6312\, ratio by using equation 3 proposed in P12. We note that even in the case of robust detections of the [O\,{\textsc{ii}}]\,$\lambda\lambda$\,7320,30\, doublet we did not make use of these measurements because of the uncertainly expected in the resulting oxygen abundances \citep[larger than t$_{2}$-t$_{3}$ relation, see][and P12]{kenn03}. There are also some \ion{H}{ii} regions in our compilation for which it is possible to obtain different measurements for the electron temperature (from both [O\,{\textsc{iii}}]\,$\lambda$\,4363\, and  [N\,{\textsc{ii}}]\,$\lambda$\,5755). In these cases we determine different abundance values for each method/auroral line as it was done in P12. Finally, in Table 3 the new results derived for the CALIFA-T$_{e}$ sample are listed. The errors in the T$_{e}$-based oxygen abundances were obtained using Monte Carlo (MC) simulations taking into account the errors on the different line ratios involved in their determination. We assumed a Gaussian distribution for the errors and that the errors among the different line ratios were not correlated for the MC. This can be considered as an upper limit for the errors since some of the contributors to these errors (e.g. uncertainties in the stellar continuum subtraction) might indeed partly be correlated. As sanity check, we also derive the electron temperature of our CALIFA-T$_{e}$ sample using {\it PyNeb} \citep[][]{pyneb}, a python tool based on the n-level atom model able to solve the equilibrium equations and obtain the physical conditions of the nebulae from the emission-line ratios. Small differences ($<$7\% in individual measurements) are found, being compatible with our error estimations.

On the other hand, we have compared the results obtained from different empirical calibrations, like the ON-calibration, the ONS-calibration (P10-ON and P10-ONS), the C-method (P12), the N2O2 method \citep{2002ApJS..142...35K}, and the method described in \cite{pena12} (hereafter PG12). The details of this comparison are beyond the scope of this article and will be presented elsewhere. For the 3423 \ion{H}{ii} regions provided by CALIFA we found most appropriate the P10-ONS calibrator, which was already anchored to oxygen abundances derived using electron temperatures. In this regard, we only summarize some important aspects of the comparison made between the different calibrations. As a summary, the aforementioned comparison concludes that there is a good correlation between the different methods once the following considerations are taken into account: (1) The P10-ON and the C-method abundances are well correlated with P10-ONS in the entire abundance range, (2) the PG12 calibration (as the authors themselves suggest), overestimates the abundance values by $\sim$0.2 dex, due to their suggested treatment on the temperature and ionizing structure of the nebula and the depletion of oxygen onto dust grains, (3) the N2O2 method is appropriate for 12+log(O/H) $>$8.6 but overestimates the abundance by 0.2-0.3 dex as it is based on a combination of photoionization and stellar population synthesis models. Thus, in the case of our CALIFA \ion{H}{ii} regions we populate the abundance sequence using the P10-ONS method (hereafter CALIFA-ONS). It correlates best with the T$_{e}$-based measurements in the whole range of metallicities, showing a good agreement with the other methods in the range of applicability of the latter and yields less outliers. It is important to emphasize here that our indirect approach, based on strong lines of oxygen, nitrogen, and sulfur (for most of the abundance range covered in this article, the ONS-calibration is similar to the R$_{23}$ calibration regarding its information content\footnote{R$_{23}$=([O\,{\textsc{ii}}]\,$\lambda\lambda$\,3727+3729 + [O\,{\textsc{iii}}]\,$\lambda\lambda$\,4959+5007)/ $\mathrm{H}\beta$, \citep[][]{1979MNRAS.189...95P}.}) is particularly important because of the wide range of ionization conditions covered by the 3423 \ion{H}{ii} regions in CALIFA (in terms of the N/O relative abundances, ionization conditions and the electron densities) compared to the regions where auroral lines are detected. 

The compilation of \ion{H}{ii} regions with electron temperatures and strong emission lines provides a more straight-forward calibration, but without the statistical significance, due to its more reduced coverage of the space of physical parameters, poorer statistics, and inhomogeneity compared to the CALIFA catalog. In that regard, for the 16 T$_{e}$-based CALIFA regions we made use of the P10-T$_{e}$ method that was based on the derivation of  t$_{3}$ temperature from the ratio of [O\,{\textsc{iii}}] auroral to nebular lines, as in the case of the oxygen abundance measurements included in the P12 dataset. 

In the following sections we want to investigate the behavior of the widely used single-parameter indicators (O3N2 and N2) with the oxygen abundance computed both from CALIFA-ONS and T$_{e}$-based measurements.

\subsection{O3N2 index}

The O3N2 index depends on two (strong) emission line ratios and was firstly introduced by \cite{alloin} as:
\begin{equation}
    \mathrm{O3N2}= \mathrm{log(\frac{\mathrm{[O~III]}\lambda5007}{\mathrm{H}\beta} \times \frac{\mathrm{H}\alpha}{\mathrm{[N~II]}\lambda 6583})}
 \end{equation}
Note that although O3N2 formally includes two line ratios in its definition, given the way the [O\,{\textsc{iii}}]$\lambda$5007 and [N\,{\textsc{ii}}]$\lambda$6584 fluxes are corrected for dust attenuation, this index is only sensitive to the extinction-corrected [O\,{\textsc{iii}}]$\lambda$5007/[N\,{\textsc{ii}}]$\lambda$6584\ ratio. Two of the most popular calibrations (among many) that relates the oxygen abundance and the O3N2 index with a simple linear regression are the ones proposed by PP04 where: 12 + log(O/H) = 8.73 - 0.32 $\times$ O3N2 and by PMC09 where: 12 + log(O/H) = 8.74 - 0.31 $\times$ O3N2. Here we also explore the calibration obtained by NAG06 using the third-order polynomial function presented in their Table 6. It is worth emphasizing that this oxygen abundance estimator is widely used in the high-metallicity regime up to the solar value, where the N2 index saturates.
In Fig$.$~\ref{o3n2_te} we show the variation of the oxygen abundance with the O3N2 index for 603 \ion{H}{ii} calibrating regions. The sample of T$_{e}$-based \ion{H}{ii} regions is represented with different symbols and colors (see the caption for details) and the 16 CALIFA \ion{H}{ii} regions are plotted with filled stars with the respective errors. All these data span a range of metallicity 7.0 $<$ 12 + log(O/H) $<$ 8.9. We also plot the PP04 calibration with a black solid line, the PMC09 calibration with a grey dashed line and the NAG06 one is plotted with a grey 3dot-dashed line. We do a robust fit (least absolute deviation method) to our T$_{e}$-based \ion{H}{ii} regions data in the range of O3N2 between $-$1.1 and 1.7 in O3N2 (green vertical dashed lines) and we find that:

\begin{equation}
12 + \mathrm{log(O/H)} = 8.533[\pm0.012] - 0.214[\pm0.012] \times \mathrm{O3N2},
 \end{equation}
 
The limits in the fit range used is mainly due to the large dispersion of the T$_{e}$-based data in the low metallicity regime (O3N2 $>$ 1.8). Hence our new empirical calibration is found by using 309 (of 603) T$_{e}$-based \ion{H}{ii} regions and in Fig$.$~\ref{o3n2_califa} is shown by the green solid line with 95 per cent (68 per cent) of the measurements within $\pm$ 0.36 dex ($\pm$ 0.18 dex). The standard errors in the zero-point and in the slope are reported within squared brackets. For all the robust fits presented in this work, the errors were computed with 10$^{5}$ bootstrap repetitions.
In Fig$.$~\ref{o3n2_califa} the CALIFA-ONS \ion{H}{ii} regions are drawn as a blue density contour-plot and also in this case we compute an independent robust fit in the range of $-$1.1 $<$ O3N2 $<$ 1.7 to the 3423 CALIFA-ONS \ion{H}{ii} regions that yields: 12 + $\mathrm{log(O/H)}$ = 8.505[$\pm$0.001] - 0.221[$\pm$0.004]$\times$ $\mathrm{O3N2}$ with $\sigma$=0.08. This fit is shown as a solid red line in Fig$.$~\ref{o3n2_califa}. The large number of regions used in deriving this latter calibration allow us to also compute the prediction intervals for each given O3N2 value, which are shown as dot-dashed red lines in this same figure ($\pm$2\,$\sigma$ in this case). These intervals can be used to derive the random error in 12+log(O/H) when this is computed from the O3N2 index. Note that the since for the ONS-based abundances we are using more line ratios than for the O3N2 calibration these prediction intervals allow us to provide an estimate of the degeneracies in the O3N2 calibration for predicting the actual oxygen abundance.
In that regard and until a larger number of precise high-metallicity T$_{e}$-based measurements are available, so the metallicity in the high regime can be derived by {\it direct} methods, the prediction intervals shown in Fig$.$~\ref{o3n2_califa} at O3N2 values below $-$0.2 should be taken with caution. Note that at these high metallicities is where the [N\,{\textsc{ii}}]$\lambda$6584\,/H$\alpha$ ratio saturates reducing the difference in information content between the ONS calibration and that given here for the O3N2 single-line ratio. 

Finally, and although the ONS-P10 calibration of the oxygen abundance is anchored to T$_{e}$-based measurements (see P10) an average offset of $\sim$0.02 dex is still found between the two calibrations (T$_{e}$-based and ONS-based). This offset should be considered as a systematic error to be added to the random error above in cases where abundance measurements based on different indices and calibrations (e.g. from galaxies at different redshifts and/or of different data quality) are to be combined. Thanks to our unprecedented statistics the equations given above represent the most update version for the O3N2 calibration. We thus confirm that our O3N2 index calibration is significantly more robust than PP04, NAG06 and PMC09 in the high abundance regime. We find a good correlation between the T$_{e}$-based and the CALIFA-ONS based measurements except for a small offset that can be due in part to the still quite low number of objects (not exactly the same as the ones used by P10) with T$_{e}$ measurements in a broad range of oxygen abundances. 

From an operational point of view, we recommend potential users of the O3N2 index to use the T$_{e}$-based calibration given above when oxygen abundances from different methods are to be combined. If only O3N2 data are to be analyzed (within a galaxy or potential changes from galaxy to galaxy) the calibration and prediction intervals based on the ONS calibration could be used instead, except for very low values of O3N2 ($<$$-$0.2) where an average T$_{e}$-based uncertainty of 0.18 dex should more reliable as in this regime the ONS abundances are heavily weighted by the value of the O3N2 index.

\begin{figure*}[t]
   \centering
  \includegraphics[scale=1.0]{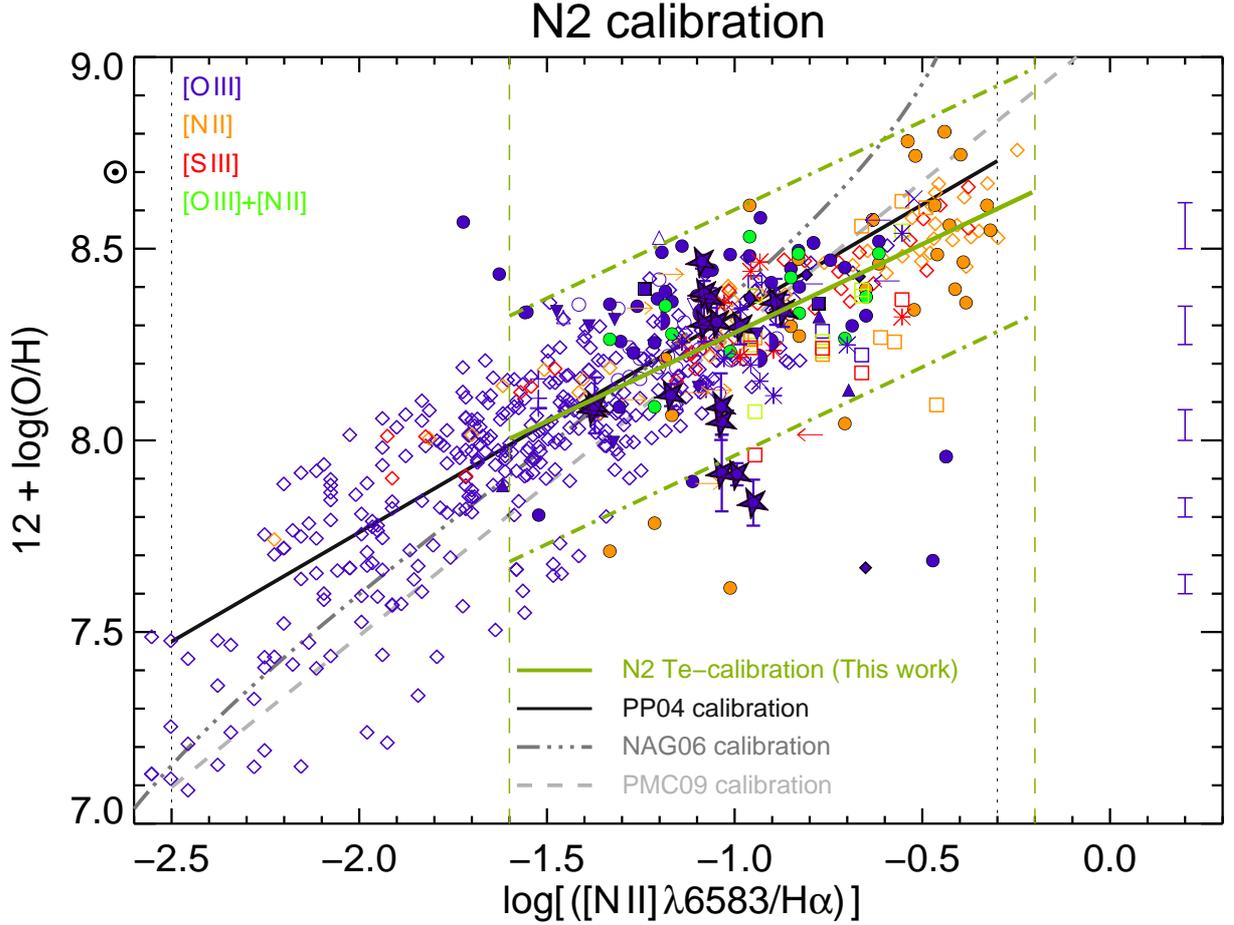}
   \caption{Oxygen abundance versus the {\it N2} index for the T$_{e}$-based \ion{H}{ii} regions. The symbols and the color code are the same as for Fig$.$~\ref{o3n2_te}. The PP04 calibration is plotted with a black solid line with its applicability interval (from N2=$-$2.5 to $-$0.3) shown with vertical grey dotted lines. The PMC09 calibration is represented with a grey dashed line while the grey 3dot-dashed line shows the NAG06 calibration. The new T$_{e}$-based calibration, 12 + log(O/H) = 8.743 +0.462$\times$N2, is showed with a green solid line while the green dot-dashed lines encompass 2\,$\sigma$ ($\pm$0.32 dex) of the measurements. The vertical dashed lines indicate the interval of our fit, from N2=$-$1.6 to$-$0.2. We only plot in blue the errors associated to the new measurements of the CALIFA-T$_{e}$ \ion{H}{ii} regions. For the sake of clarity we show in the right the average random error associated to the computation of the oxygen abundances for different metallicity bins.}
              \label{n2_te}%
    \end{figure*}
    
    \begin{figure*}[t]
   \centering
  \includegraphics[scale=1.0]{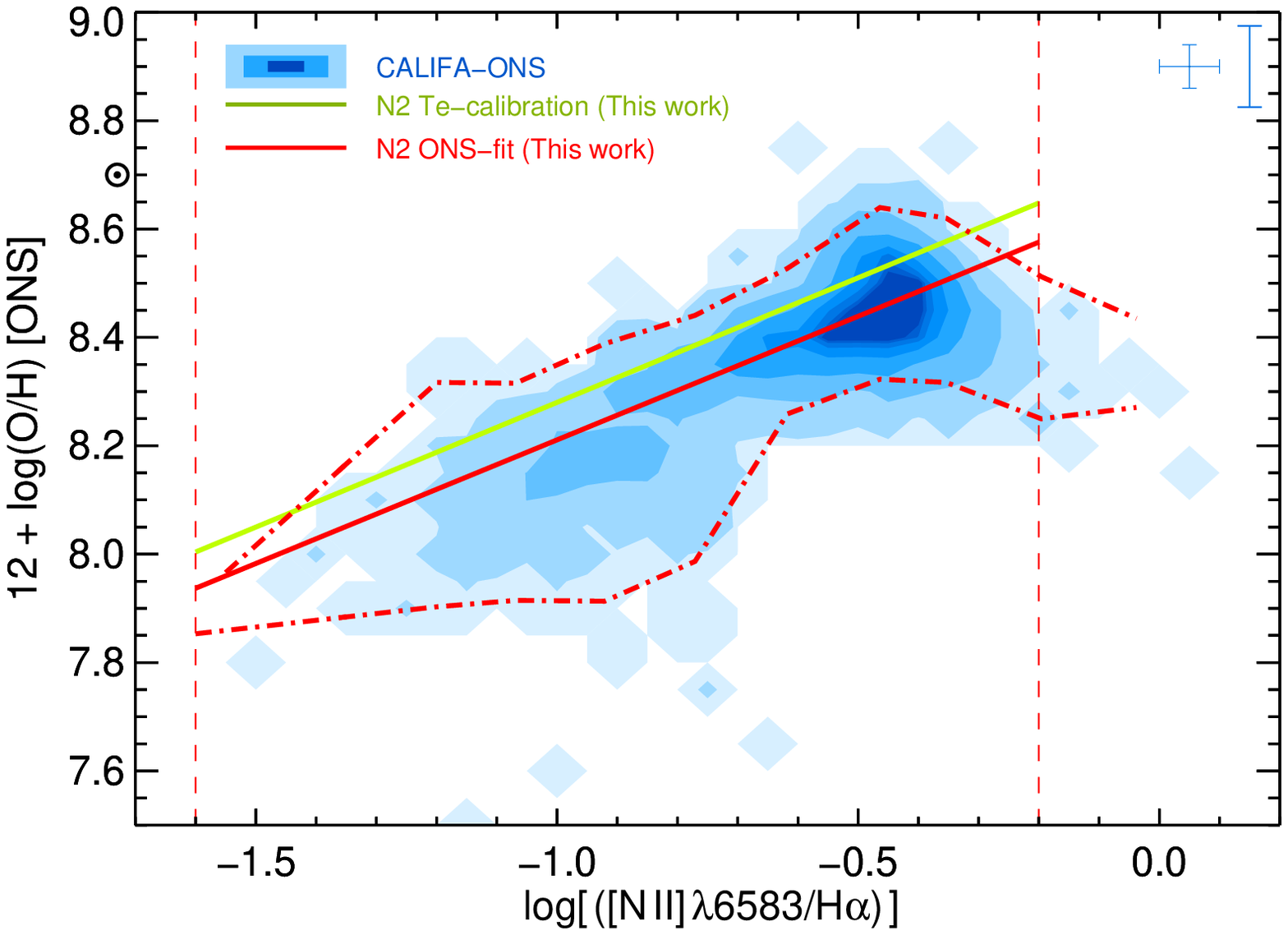}
   \caption{Oxygen abundance versus the {\it N2} index for the \ion{H}{ii} complexes within CALIFA galaxies. We plot the CALIFA-ONS \ion{H}{ii} regions as a blue density contour-plot. Our new N2 calibration (see Fig$.$4) is plotted with a green solid line. Red lines are used to show the CALIFA-ONS best linear fit (solid) and its 2\,$\sigma$ prediction intervals (dot-dashed) with its applicability interval (from N2=$-$1.6 to $-$0.2, vertical dashed lines) (see text for details on these fits). In this case we obtain 12 + log(O/H) = 8.667 +0.455$\times$N2. The blue cross at the top right corner of the plot indicate the average random (left) of the CALIFA-ONS data points and the typical systematic (right) error associated to the measurement of ionized-gas oxygen abundances.}
              \label{n2_califa}%
    \end{figure*}
    
\subsection{N2 index}

The N2 index was studied by several groups including \cite{storchi}, \cite{vanzee}, \cite{raimann}, \cite{denic} and it is defined as:
\begin{equation}
\mathrm{N2}= \mathrm{log(\frac{\mathrm{[N~II]}\lambda 6583}{\mathrm{H}\alpha})}
\end{equation}

Despite the saturation that the N2 suffer in the high-metallicity regime, this parameter is a very useful indicator to calculate the oxygen abundance for different reasons: (1) it is very sensitive to the metal content of a nebulae; (2) does not suffer of reddening correction or flux calibration issues due to the close wavelength of these lines; and (3) can be detected with new generation near-infrared spectrographs in 8-10m class telescopes at high redshifts. For the N2 index PP04 and PMC09 proposed the following relations: 12 + log(O/H) = 8.90 + 0.57 $\times$ N2 and 12 + log(O/H) = 9.07 + 0.79 $\times$ N2, respectively. Our results along with the PP04, NAG06 and PMC09 calibrations are shown in Fig$.$~\ref{n2_te}. The symbols and colors are the same as in Fig$.$~\ref{o3n2_te}. The robust regression made to 452 T$_{e}$-based abundance measurements gives:

\begin{equation}
12 + \mathrm{log(O/H)} = 8.743[\pm0.027] + 0.462[\pm0.024] \times \mathrm{N2},
 \end{equation}

\noindent
with 95 per cent (68 per cent) of the measurements of log(O/H) found within $\pm$0.32 dex ($\pm$ 0.16 dex), in the interval of $-$1.6 $<$ N2 $<$ $-$0.2. On the other hand, the robust fit to the CALIFA-ONS \ion{H}{ii} regions yields to 12 + $\mathrm{log(O/H)}$ = 8.667[$\pm$0.006] + [0.455$\pm$0.011] $\times$ $\mathrm{N2}$ with an average rms of $\sigma$=0.09. As in the case of the O3N2 calibrations, these new calibrations provide somewhat shallower slopes than the PP04, NAG06 and PMC09 calibrations. As expected, this calibrator has a monotonic behavior with the oxygen abundance but saturates at log([N\,{\textsc{ii}}]$\lambda$6584/H$\alpha$)$\sim$$-$0.4 where the piling of both T$_{e}$-based and CALIFA-ONS \ion{H}{ii} regions is evident. We also find a good agreement between the trend of the T$_{e}$-based \ion{H}{ii} regions and the CALIFA-ONS \ion{H}{ii} regions although there is a systematic offset of 0.08 dex (in the sense that T$_{e}$-based abundances are slightly larger than ONS-based ones) that should be taken into account should the ONS-based calibration be used in combination with other oxygen abundance estimates. The strength of the large collection of CALIFA data is clear in this case, where the prediction intervals can be safely adopted within the entire range of applicability of the N2 index given the comparatively small dependence of the ONS calibration on the value of the N2 line ratio itself, especially as we move to high metallicities where the latter saturates.

\section{Discussion and Conclusions}
In this paper we provide revisited empirical calibrations for the oxygen abundances in \ion{H}{ii} regions based on the O3N2 and N2 indicators. This work represents the most comprehensive compilation of both T$_{e}$-based and multiple strong-line (ONS-based) ionized-gas abundance measurements in external galaxies to date.

The differences found between these calibrations and those obtained by PP04, NAG06 and PMC09 recommend revising the results obtained using the latter calibration. These differences are particularly noticeable at the high metallicity regime (12+log(O/H) $>$ 8.2) and on one hand they are mainly due to the lack of high quality observations of \ion{H}{ii} regions with auroral lines at the high-metallicity end at the time of the publication of the PP04 calibration, and thus their justified need for including predictions from photoionization models at this regime. On the other hand, the large differences found at high metallicity between our calibration and the one of NAG06 are due to the fact that NAG06 in this metallicity range is using a sample of galaxies (their sample C, for more details see Sec$.$2.2 in \cite{nag06}) that have no T$_{e}$-based abundances measured.
We find that a linear relation provides a good fit to the oxygen abundance as a function of the O3N2 (N2) parameters with rms values of 0.18 dex (0.16 dex). The relatively low statistics and rather heterogeneous origin of the direct abundance measurements precludes deriving reliable predicting intervals as a function of these parameters or the derived oxygen abundance. This is where the results of the analysis of large and homogeneous set of ONS-based measurements are most useful with the caveats and limitations described in Sect$.$ 3$.$1 and 3$.$2. 

Hereafter we discuss on the potential physical origin of the relations shown in Figures 2 through 5. The ionization degree of the \ion{H}{ii} region also plays an important role when deriving a proper oxygen abundance. Many other empirical methods based on both observations \citep[e.g$.$][]{p01a, p01b, pilythuan} and photoionization models \citep[e.g$.$][]{mcgaugh, 2002ApJS..142...35K} do consider an ionization parameter --typically the [O\,{\textsc{iii}}]/[O\,{\textsc{ii}}] ratio-- to improve the accuracy of their derived O/H values. Indeed, it has been shown that empirical calibrations which do not consider an ionization parameter have a larger scatter (which is even larger than observational errors) than those ones that do not use it \citep[][]{kob99, angel10, moustakas10,angel12}. That is particularly important when dealing with IFS data, as observations may be actually showing the ionization structure of the \ion{H}{ii} regions \citep[e.g$.$][]{carol08,angel11, ana10, ana11, ana12, epm11, james13a, james13b}. This behavior is clearly seen in Fig$.$ 13 of LS11, which compares the [O\,{\textsc{iii}}]/[O\,{\textsc{ii}}] ratio and oxygen abundance as provided by several empirical calibrations. Using N2 and O3N2 calibrations proposed by PP04, NAG06 and PMC09 it is clearly seen that regions with higher ionization degree tend to have lower oxygen abundances (see also Fig$.$ 12 of PP04). Another important issue to consider is that the ionization parameter is also related to the age of the most-recent star formation event. Galaxies hosting young starbursts will have higher ionization parameters than galaxies where the main star-formation event happened long time ago. This fact introduces an extra bias to all O/H estimations based on calibrations which do not consider an ionization parameter for the gas \citep[][]{sta10, angel10}. In the case of objects with high N/O ratios \citep[e.g$.$][]{pustilnik, angel07, angel10, angel11, ana10, ana12, epm11, amorin12, carol13} the use of a [N\,{\textsc{ii}}]-based calibration will provide O/H values which are higher than the real ones. According to the photoionization models described in \cite{pmd05} and PMC09 all the dispersion found in the relation between both O3N2 and N2 parameters and the oxygen abundances derived from the T$_{e}$ method can be explained in terms of an additional dependence on ionization parameter and on nitrogen-to-oxygen ratio. In the case of our compilation, we have checked that the grid of models presented in PMC09 covers all our observed sample by varying of these two parameters. 

The comparison between the T$_{e}$- and ONS-based abundances in the case of the O3N2 yields a very small difference 0.02 dex, which can be attributed to the intrinsic dispersion in ONS for a given T$_{e}$-based abundance (see P10). In the case of N2, on the other hand, we find a systematic difference of 0.08 dex which is evident as an offset between the green and red lines of Figure 5. This cannot be due exclusively to the limitations of the ONS abundances along but mainly to differences between either the N2 or ONS measurements of the CALIFA regions compared to those where T$_{e}$-based measurements have been derived. Since for a given O3N2 we find a very good agreement between the T$_{e}$- and ONS-based measurements, it is more likely that the difference arises in the N2 measurements (only 16 CALIFA T$_{e}$-based regions are not enough to directly confirm this). One possibility to explain the differences between the trend of the ONS-fit and our T$_{e}$-based calibration in the case of N2 is that the change in N2 (for a given O3N2 and ONS) is related to spatial-resolution effects. The reasoning behind this is that the vast majority of the T$_{e}$-based measurements come from long-slit or multi-object spectroscopy of very nearby galaxies (typically inside the Local Volume; d$<$11\,Mpc), where a 1\,arcsec resolution yields a physical resolution better than 50\,pc, while the typical physical resolution of the CALIFA data is, in the best case, four times this number. In this regard, Mast et al. (2013, in prep.) do not find a systematic increase in the N2 index with decreasing spatial resolution using simulations based on IFS data on very nearby systems from PINGS \citep{pings}. However, since their work is not based on a large diameter-limited sample like CALIFA, we cannot completely rule out this effect. On the other side, we should also consider the fact that while a significant fraction of the \ion{H}{ii} regions included in our T$_{e}$-based analysis were extracted from dwarf galaxies, these are almost absent (by design) from the CALIFA sample, which is mainly constituted by disk galaxies. Therefore, any difference (for a given oxygen abundance) in the N/O ratio or the excitation conditions of the nebulae in dwarfs with respect to disk galaxies could also result in a different N2 for the CALIFA sample even for the same O3N2 and ONS measurements. Should that be the case, future analyzes on the chemical abundances in galaxies based solely on the N2 index should pay close attention to the sample selection since systematics in the N2-based abundances could arise if galaxies of different types are combined, for example, at different redshifts.

Our new calibration has also important implications on studies investigating the gas metallicity evolution of galaxies up to high redshifts. For example, the mass-metallicity relation (MZR) at z$>$2 presented by \cite{erb}, which is often used as a reference sample, was based on the N2 calibration of PP04. They found an evolution in MZR of $\sim$0.3 dex at z$>$2 compared to the local SDSS sample. However, the metallicity evolution is enhanced using our new N2 calibration by $\sim$0.1$-$0.2 dex. Our Eq$.$4 would lead to an average decrease of 0.4 dex to z$\sim$2, while using the fit to the CALIFA-ONS \ion{H}{ii} regions results in an evolution of 0.5 dex. This is more in accordance with recent determinations of metallicity in distant zCOSMOS galaxies exploiting measurements of all 5 strong lines \citep[][]{epm13}.

Thus, we conclude that the relations given in this paper will allow improving our understanding on the chemical evolution of the Universe even when only single-parameter abundance measurements are available and as long as the predicting intervals derived here are taking into account. Possible effects associated to spatial resolution and sample selection should also adequately be accounted for.

\begin{acknowledgements}
We would like to thank the anonymous referee for the review performed to the manuscript. His/her comments and suggestions helped to improve the content of the paper. 
R.A. Marino is funded by the Spanish program of International Campus of Excellence Moncloa (CEI). This study makes uses of the data provided by the Calar Alto Legacy Integral Field Area (CALIFA) survey (http://www.califa.caha.es). Based on observations collected at the Centro Astron\'omico Hispano Alem\'an (CAHA) at Calar Alto, operated jointly by the Max-Planck-Institut f\"ur Astronomie and the Instituto de Astrof\'\i sica de Andaluc\'\i a (CSIC).
CALIFA is the first legacy survey being performed at Calar Alto.The CALIFA collaboration would like to thank the IAA-CSIC and MPIA-MPG as major partners of the observatory, and CAHA itself, for the unique access to telescope time and support in manpower and infrastructures.  The CALIFA collaboration thanks also the CAHA staff for the dedication to this project. We thank the {\it Viabilidad , Dise\~no , Acceso y Mejora } funding program, ICTS-2009-10, for supporting the initial developement of this project.
S.F.S., F.F.R.O. and D. Mast thank the {\it Plan Nacional de Investigaci\'on y Desarrollo} funding programs, AYA2012-31935 of the Spanish {\it Ministerio de Econom\'\i a y Competitividad}, for the support given to this project. S.F.S thanks the the {\it Ram\'on y Cajal} project RyC-2011-07590 of the spanish {\it Ministerio de Econom\'\i a y Competitividad}, for the support giving to this project. F.F.R.O. acknowledges the Mexican National Council for Science and Technology (CONACYT) for financial support under the programme Estancias Posdoctorales y Sab\'{a}ticas al Extranjero para la Consolidaci\'{o}n de Grupos de Investigaci\'{o}n, 2010-2012. We acknowledge financial support for the ESTALLIDOS collaboration by the Spanish Ministerio de Ciencia e Innovaci\'{o}n under grant  AYA2010- 21887-C04-03. BG-L also thanks the support from the Spanish {\it Ministerio de Econom\'\i a y Competitividad} (MINECO) under grant AYA2012- 39408-C02-02. J.~F.-B. acknowledges financial support from the Ram\'{o}n y Cajal Program and grant AYA2010-21322-C03-02 from the Spanish Ministry of Economy and Competitiveness (MINECO), as well as to the DAGAL network from the People Programme (Marie Curie Actions) of the European Union's Seventh Framework Programme FP7/2007-2013/ under REA grant agreement number PITN-GA-2011-289313. CK has been funded by the project AYA2010-21887 from the Spanish PNAYA. P.P. acknowledges support by the Funda\c{c}\~{a}o para a Ci\^{e}ncia e a Tecnologia (FCT)
under project FCOMP-01-0124-FEDER-029170 (Reference FCT PTDC/FIS-AST/3214/2012), funded by FCT-MEC (PIDDAC) and FEDER (COMPETE). R.M.G.D. and R.G.B. also thanks the support from the Spanish {\it Ministerio de Econom\'\i a y Competitividad} (MINECO) under grant AyA2010-15081. V.S., L.G. and A.M.M. acknowledge financial support from Funda\c{c}\~{a}o para a Ci\^{e}ncia e a Tecnologia (FCT) under program Ci\^{e}ncia 2008 and the research grant PTDC/CTE-AST/112582/2009.
\end{acknowledgements}

\bibliographystyle{aa}
 
\bibliography{referencias}

\end{document}